\documentstyle[12pt,epsfig]{article}

\def\beq{\begin{equation}}
\def\eeq{\end{equation}}

\def\beqa{\begin{eqnarray}}
\def\eeqan{\end{eqnarray}}

\def\Title#1{\begin{center} {\Large #1 } \end{center}}

\begin{document}

\Title{ Rare $\Phi$ Decays and Exotic Hadrons }

\bigskip\bigskip


\begin{raggedright}  

{\it S.I.Serednyakov\index{Serednyakov, S.I.}\\
Budker Institute of Nuclear Physics 
\\
and Novosibirsk State University,\\
Novosibirsk, 630090, Russia }
\bigskip\bigskip
\end{raggedright}

\section{Introduction}
Exotic hadrons are named those hadrons, which structure is different from ordinary $q_1\bar{q_2}$
 structure for mesons and $q_1q_2q_3$ for baryons.  Their exotic nature could  reveal itself in unusual
 properties of these hadrons like some suppressed or enhanced  decays,  too wide or too narrow  widths,
 quantum numbers, forbidden in conventional structure, etc.   Up to now among huge variety of hadrons
 about 10 candidates were found, which look like exotic states. In scalar meson sector such
 candidates are  the lowest lying states $f_0(980)$ and $a_0(980)$. The main reasons, which lead to the
 conclusion on possible exotic structure of  $f_0(980)$ and $a_0(980)$,  are their suppressed production
 in $J/\Psi$ decays, low values  of $\gamma\gamma$-width,  too low masses.  

      Three models are used  to describe $f_0$ and $a_0$ mesons \cite{Modl}
- conventional $q\bar{q}$ model, molecular model ($K\bar{K}$) and 4-quark model ($q\bar{q}q\bar{q}$).
 The $q\bar{q}$ model is hardly consistent with experimental data.  More than 10 years ago the radiative
 decays $\phi\to f_0\gamma ,a_0\gamma$ were proposed as a new sensitive test of $f_0$ and  
$a_0$ structure \cite{Pred}.  

     These decays were studied recently in the  reactions:
\beq   e^+e^-\to\phi\to\pi^0\pi^0\gamma , \pi^+\pi^-\gamma  \label{ppg0} \eeq
\beq   e^+e^-\to\phi\to\eta\pi^0\gamma   \label{etpg0} \eeq
which could proceed via radiative decay $\phi (1020)\to f_0\gamma ,a_0\gamma$. We measured the branching
 ratios of these decays  and of other rare decays of $\phi$ and $\rho (770)$, $\omega (783)$.
 Other experimental data from Novosibirsk and conference contribution from IHEP, 
Protvino  are reviewed in this talk.

\section{Experiment}  The experiments to study the reactions (\ref{ppg0}),  (\ref{etpg0}) 
have been carried out at VEPP-2M collider in the  energy range  2E  from 0.4 to 1.4 GeV.
  VEPP-2M is the lowest energy $e^+e^-$ collider, operating in Novosibirsk since 1974. The collider
 luminosity $L$ sharply depends on its  energy $L\sim E^4$.  At the energy of $2E=M_{\phi}$ the maximum 
 luminosity  $L_{max}=5\cdot 10^{30}cm^{-2}s^{-1}$.  
  
\begin{figure}[ht]
\begin{minipage}[ht] {0.98\textwidth}
\epsfig{figure=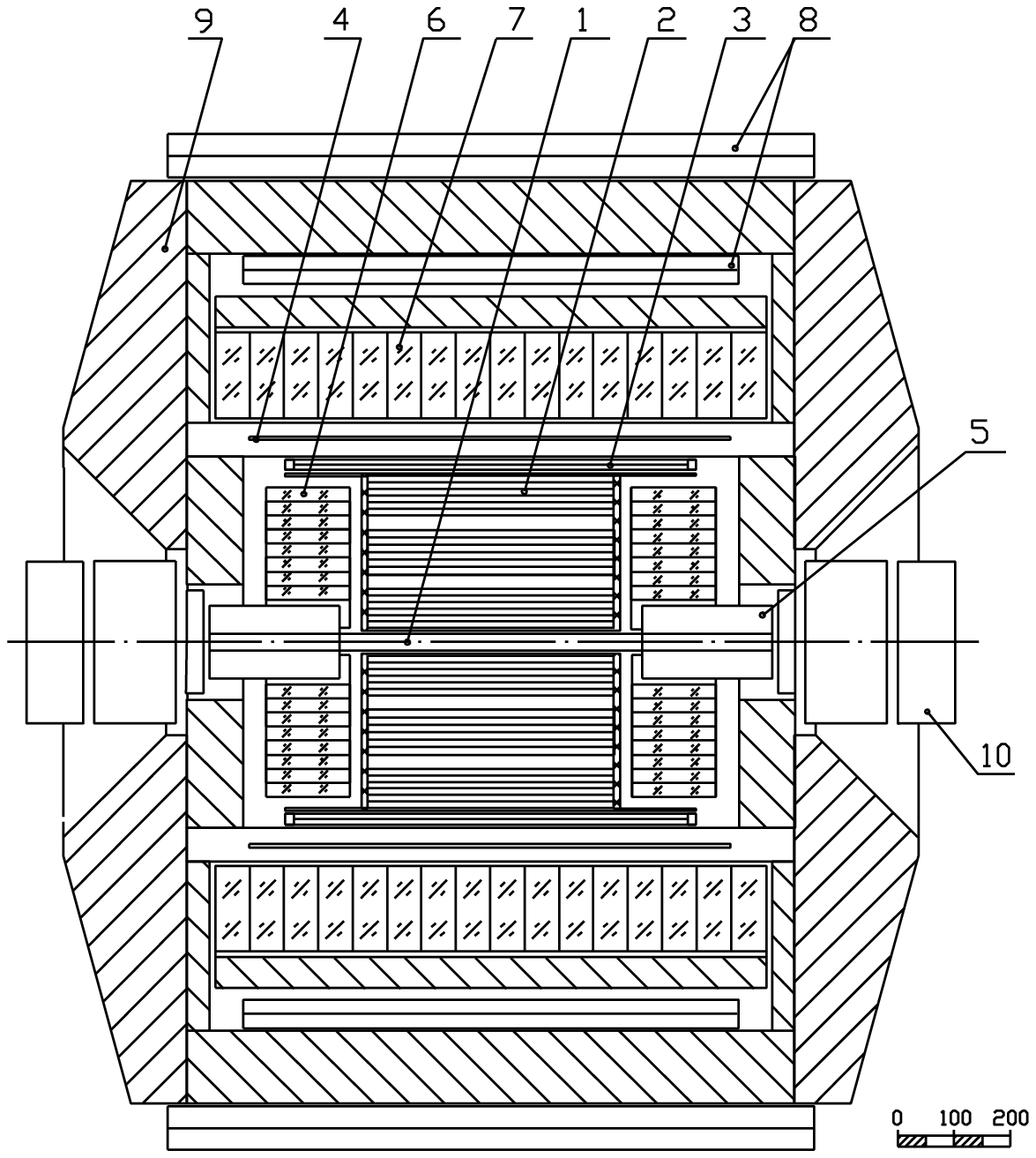,width=0.48\textwidth}
\epsfig{figure=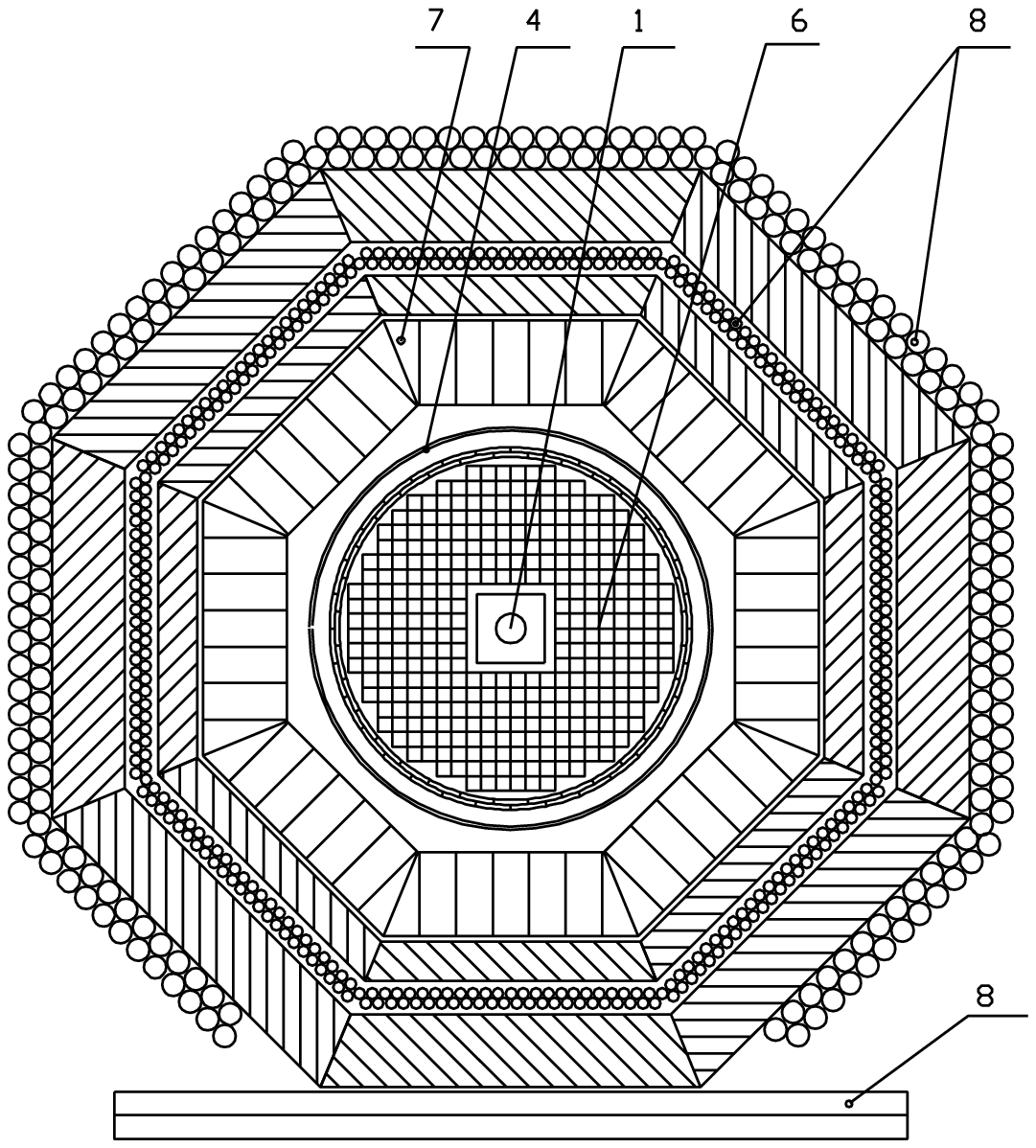,width=0.48\textwidth}
\end {minipage}
\caption{Layout of CMD-2 detector; 1 - beam pipe;  2 - drift chamber; 3 - Z-chamber;
 4 - superconductive solenoid;   5 - compensating magnet; 6 - endcap BGO calorimeter;
  7 - barrel CsI(Tl)calorimeter;  8 - muon range system; 9 - yoke;   10 - quadrupole lenses}
\label{CMD2}
\end{figure}
     At present two  detectors CMD-2 and SND, located opposite each other,  take data.
  CMD-2 \cite{CMD}  is a magnetic  detector  (fig.\ref{CMD2}) with superconductive solenoid
 and  20 layer drift chamber with jet  cell structure. Electromagnetic calorimeter consists
 of 892  CsI(Tl) crystals in barrel and of 680 BGO crystals in endcaps. The muon
 identification is provided by 4 layers of streamer tubes inside the yoke. The CMD-2
 detectors operates at VEPP-2M since 1992 with $\simeq 27 pb^{-1}$ of collected luminosity. 

\begin{figure}[htb]
\begin{minipage}[htb] {0.98\textwidth}
\epsfig{figure=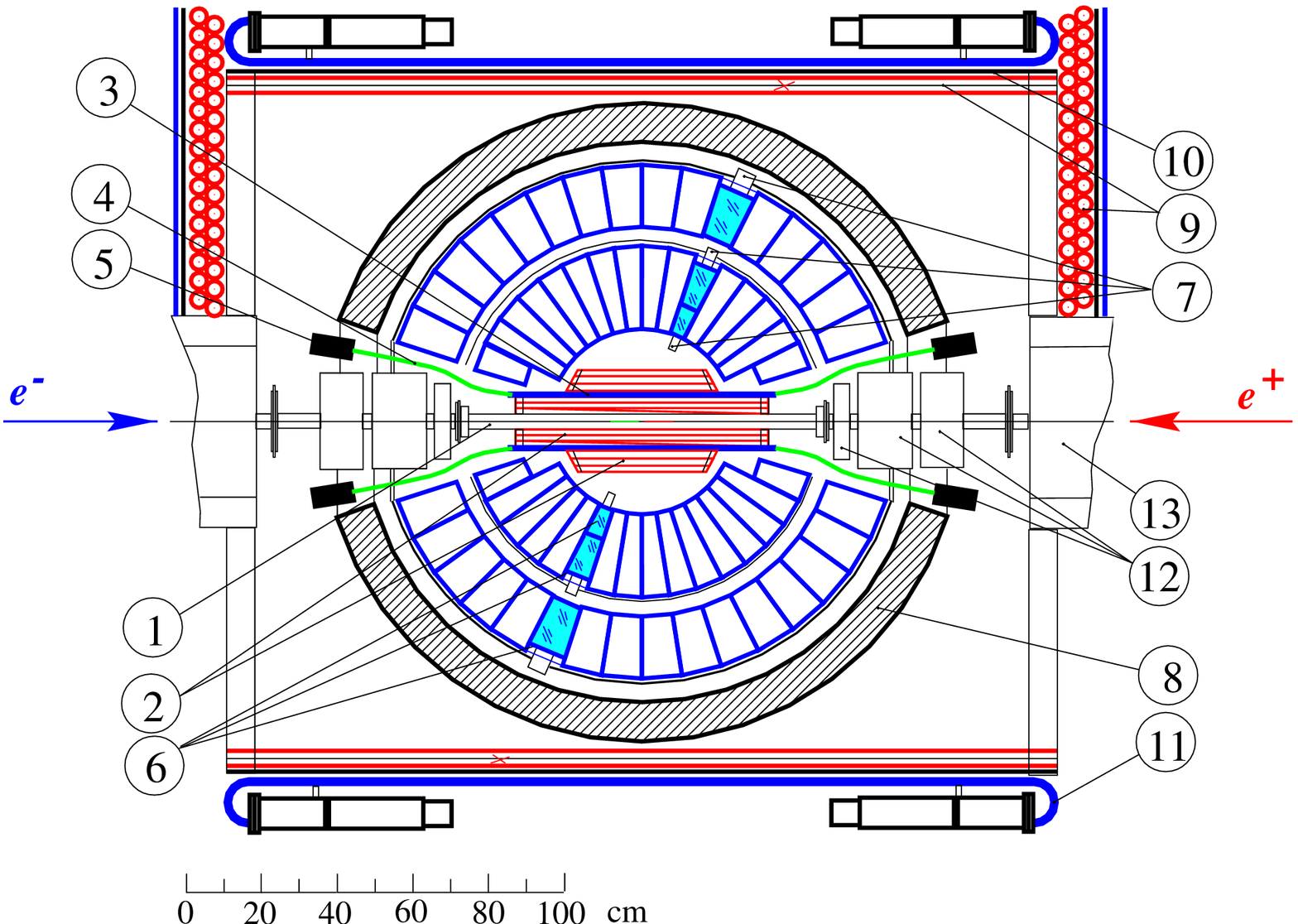,width=0.48\textwidth}
\epsfig{figure=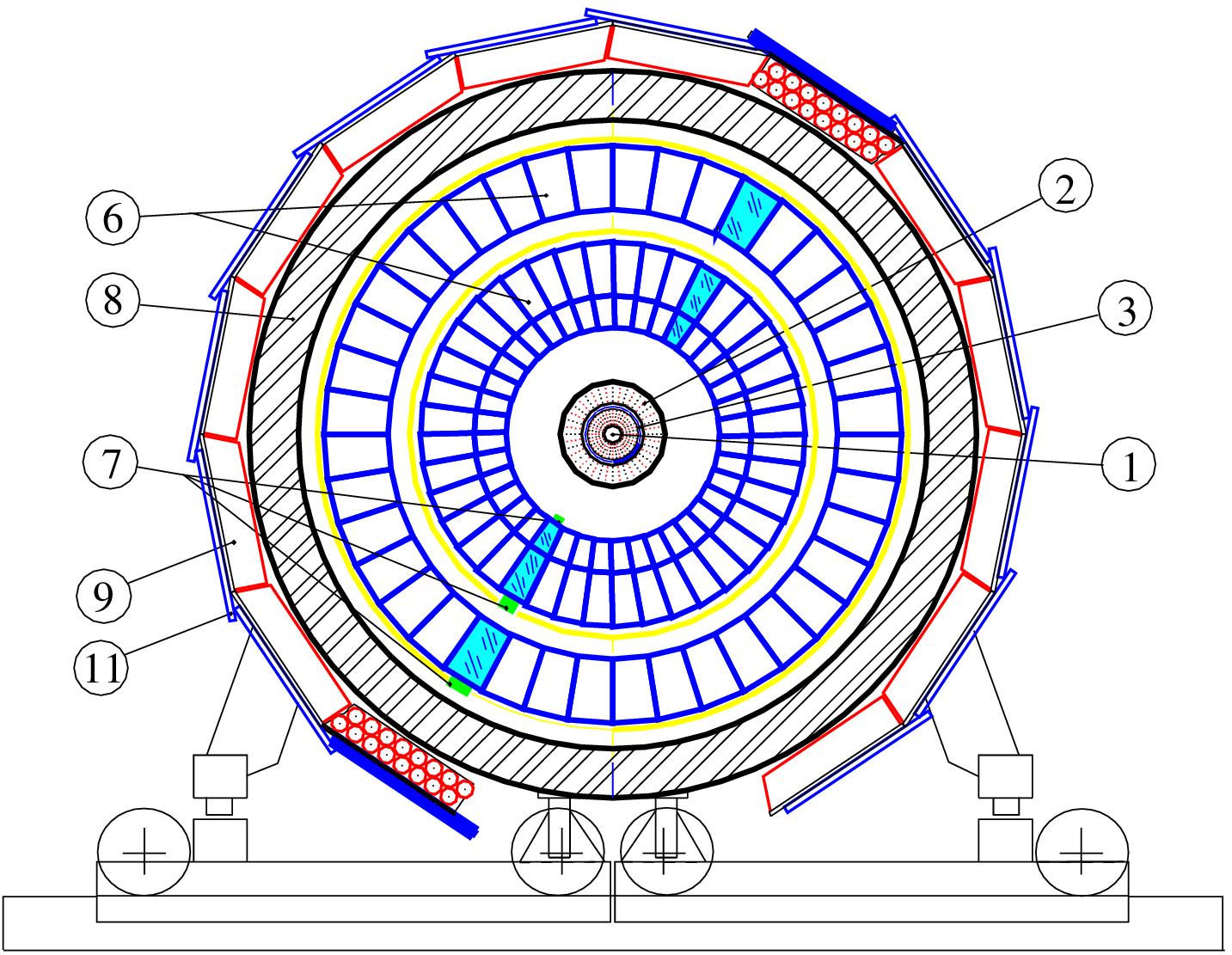,width=0.48\textwidth}
\end {minipage}
\caption{ Layout of SND detector; 1 - beam pipe, 2 - drift chambers,
 3 - scintillation counter, 4 - light guides, 5 - PMTs,  6 - NaI(Tl)
 crystals,  7 - vacuum phototriodes, 8 - iron absorber, 9 - streamer tubes,
  10 - iron plates,  11 - plastic scintillators, 12 and 13 - collider magnets}
\label{SND}
\end{figure}

     SND   \cite{snd} is a general purpose nonmagnetic detector   (fig.\ref{SND}). The main
 part of SND is three-layer spherical electromagnetic calorimeter with 1625 NaI(Tl) crystals
 of 3.6 t total weight. Detector includes also a 10-layer drift chamber
 and outer muon system, consisting of streamer tubes and plastic scintillation counters.  SND
 resembles famous Crystal Ball detector constructed in SLAC, but unlike Crystal Ball it has a 3-layer
 crystal calorimeter, which  provides better particle recognition  $e/\pi /\mu$ and $\gamma /K_L$.
  The integrated luminosity accumulated by SND since 1995 is about $27 pb^{-1}$.

    Both detectors take data in parallel. Total number of produced resonances  is
  $N_{\phi}\simeq 4.5\cdot 10^7$,  $N_{\rho}\simeq 4\cdot 10^6$,   $N_{\omega}\simeq 2.5\cdot 10^6$.
About half of the total time was used for scanning the  energy range between resonances
 with the goal of the  precise measurement of the quantity
 $R=\frac{\sigma(e^+e^-\to hadrons)}{\sigma(e^+e^-\to\mu^+\mu^-)}$
 and study  particular channels of $e^+e^-$-annihilation.

\section{Evidence of the decays $\phi\to f_0\gamma ,a_0\gamma$}  
 The first search for the decays
  $\phi\to f_0\gamma ,a_0\gamma\to\pi^0\pi^0\gamma , \eta\pi^0\gamma$ 
 was carried out with ND detector \cite{ND} at VEPP-2M collider in 1987 
\cite{ZShr, NData}.   In that early experiment the upper limits on the
 decays branching ratios at a level $\sim 10^{-3}$ were placed.  Later it was
 shown by N.Achasov \cite{Pred}, that study of these decays can provide a unique
 information on the structure of lightest scalars $f_0$ and $a_0$. Subsequent
 studies proved this idea. In 1995 the  experiments started at VEPP-2M with SND
 detector \cite{snd}, which has photon detection capabilities much better than ND.
  Study of the decays  $\phi\to f_0\gamma ,a_0\gamma\to\pi^0\pi^0\gamma , \eta\pi^0\gamma$
 was one of important goals of SND detector.  In 1997 the first results from SND \cite{H97}  were reported
 with evidence of the processes  (\ref{ppg0}),  (\ref{etpg0}). 

    The reaction   (\ref{ppg0}) was studied by SND in neutral final state :
\beq   e^+e^-\to\phi\to\pi^0\pi^0\gamma \label{ppg1}     \eeq
  so both processes (\ref{ppg0}) and (\ref{etpg0})  were studied in 5 photon
 final state. The main background comes from the following reactions:
\beq  
e^+e^-\to\phi\to\eta\gamma \to 3\pi^0\gamma \label{etg}     \eeq
\beq  
e^+e^-\to\omega\pi^0\to\pi^0\pi^0\gamma   \label{omp} \eeq
\beq  
e^+e^-\to K_SK_L\to neutrals  \label{kskl} \eeq

     In order to suppress background the SND events were selected with 5 photons,
 satisfying energy-momentum balance. The final state should contain $2\pi^0$  for
 the process (\ref{ppg1}) or $\eta\pi^0$ for (\ref{etpg0}).  The contribution of
 the reaction  (\ref{etg}) into the process (\ref{etpg0})  was suppressed by the
 cut on the maximum energy of the photon in an event.  For suppression of  the
 process (\ref{omp}) the cuts were imposed on the $\pi^0\gamma$ effective mass,
 excluding the  region around $\omega (783)$-mass.   The processes  (\ref{etg}) 
 and (\ref{kskl}) were suppressed by the parameter, describing the transverse shower
 profile in the calorimeter \cite{Bozh}.
 
     Under chosen selection criteria the detection efficiency  was determined  to  be $15\%$ and  $4\%$  
for the processes  (\ref{ppg1}) and  (\ref{etpg0})   respectively.  In the experimental data sample with 
integrated luminosity  $4 \  pb^-1$ about 150 events of process   (\ref{ppg1}) were found.  The number 
of found events  of  the process  (\ref{etpg0}) in the full SND  sample of $\simeq 2\cdot 10^7$ produced
 $\phi$ was $\simeq$ 70 events.  

      The angular distributions of the  processes   (\ref{etpg0}),  (\ref{ppg1})
  were studied in refs. \cite{PLppg, PLetpg}.  It was shown,  that the distribution
 over polar angle $\theta$ of the recoiled  photon is proportional to ($1+cos^2\theta$).
  The angle $\psi$ was defined as an angle between a pion direction in the $\pi^0\pi^0$ 
or   $\eta\pi^0$  center of mass reference frame and the recoiled photon direction.  The
 distribution over  $cos\psi$  was found to be flat.  So, the experimental data confirm 
the conclusion that   $\pi^0\pi^0$  and   $\eta\pi^0$ system are produced in scalar state. 

\begin{figure}[htb] 
\begin{minipage}[htb]{70 mm}
\epsfig{file=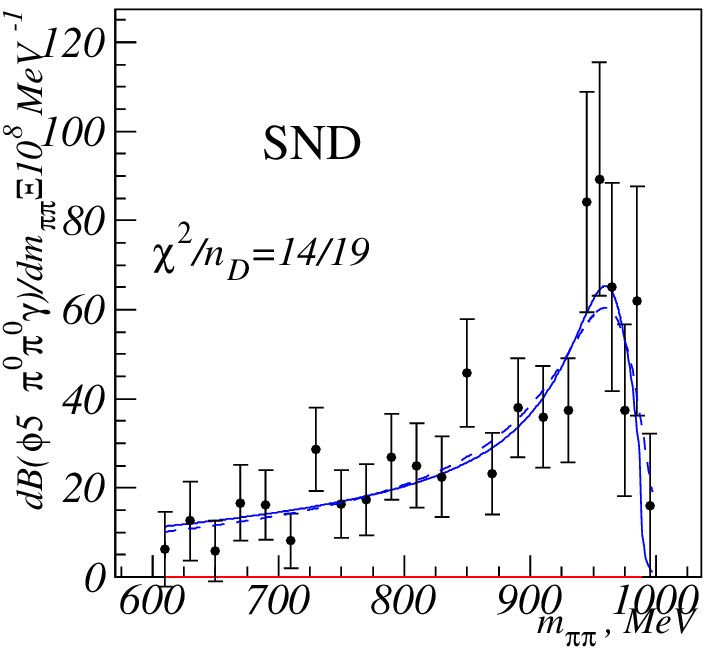, width=0.9\textwidth}
\caption{The $\pi^0\pi^0$  mass distribution in the process $ e^+e^-\to\pi^0\pi^0\gamma$ }
\label{pi0}
\end{minipage}
\begin{minipage}[htb]{70 mm}
\epsfig{file=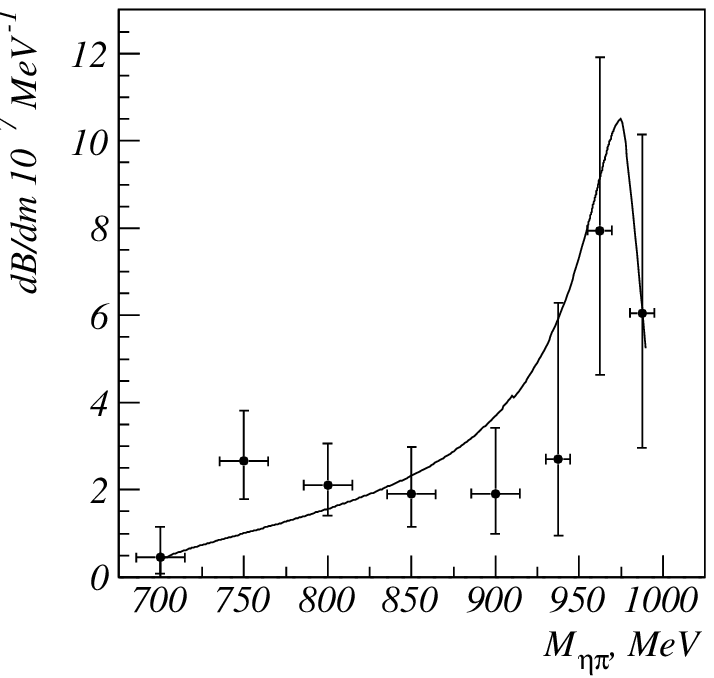, width=0.9\textwidth}
\caption{The $\eta\pi^0$  mass distribution in the process $ e^+e^-\to\eta\pi^0\gamma$}
\label{etapi}
\end{minipage}
\end{figure}

         The study the  $\pi^0\pi^0$  and $\eta\pi^0$ mass spectra    was important
 for the  interpretation of the data.  Figs. \ref{pi0},\ref{etapi} show the obtained
 mass spectra after background subtraction and detection efficiency
 corrections. Both pictures show the considerable rise in the spectra at higher masses.
 The visible location of the peak in spectra is near  960~MeV.  The table with
 numerical values of $\pi^0\pi^0$  mass spectrum can be found in \cite{PLppg}.

       Summing data from the mass spectra in figs \ref{pi0},\ref{etapi} and CMD-2 data 
one can obtain the branching ratios for particular mass ranges:
 
1 -  SND result for $m_{\pi\pi}> 900$~MeV \cite{PLppg}:
\beq B(\phi\to\pi^0\pi^0\gamma)=(0.50\pm 0.06\pm 0.06)\cdot 10^{-4} \label{pp1} \eeq 

2 -  SND result for the whole mass spectrum  \cite{PLppg}:
\beq B(\phi\to\pi^0\pi^0\gamma)=(1.14\pm 0.10\pm 0.12)\cdot 10^{-4} \label{pp2} \eeq 
here and below the first error is statistical while the second one is systematic,
 which is determined mainly by the background subtraction error,  detection 
efficiency error and normalization error.

3 -  CMD-2 result for  $m_{\pi\pi}> 700$~MeV \cite{CMD1}:
\beq B(\phi\to\pi^0\pi^0\gamma)=(0.92\pm 0.08\pm 0.06)\cdot 10^{-4} \label{pp3} \eeq 

4 -   SND result for $m_{\eta\pi^0}> 950$~MeV \cite{H99}:
\beq B(\phi\to\eta\pi^0\gamma)=(0.36\pm 0.11\pm 0.03)\cdot 10^{-4} \label{etp1} \eeq 

5 -   SND result for the whole mass spectrum \cite{H99}:
\beq B(\phi\to\eta\pi^0\gamma)=(0.87\pm 0.14\pm 0.07)\cdot 10^{-4} \label{etp2} \eeq 

6 -   CMD-2 result for the whole mass spectrum \cite{CMD1}:
\beq B(\phi\to\eta\pi^0\gamma)=(0.90\pm 0.24\pm 0.10)\cdot 10^{-4} \label{etp3} \eeq 

              All results listed above are practically model independent, because they  do
 not use  an assumption about $f_0$ or $a_0$ contributions into the final state.  Then,
 assuming  $f_0$ and  $a_0$ dominance in the final state, using relation based on
 isotopic invariance
 $B(\phi\to\pi^+\pi^-)=2B(\phi\to\pi^0\pi^0)$, and neglecting the decay $\phi\to KK\gamma$,
  we can obtain for the decay $\phi\to f^0\gamma$ and $\phi\to a^0\gamma$: 

7 -  SND result \cite{PLppg}:
\beq B(\phi\to f^0\gamma)=(3.42\pm 0.30\pm 0.36)\cdot 10^{-4} \label{f4} \eeq 

8 -  SND result \cite{H99}:
\beq B(\phi\to a^0\gamma)=(0.87\pm 0.14\pm 0.07) \cdot 10^{-4} \label{a4} \eeq 

9 -  CMD-2 result \cite{CMD1}:
\beq B(\phi\to f^0\gamma)=(2.90\pm 0.21\pm 1.54)\cdot 10^{-4} \label{f5} \eeq 

      The analysis of the $\pi^0\pi^0$ mass spectrum was done on the base of the
  work \cite{Pred}. The spectrum was described by a sum of contributions from
 $f_0$  and $\sigma$ mesons \cite{Ivanch}.  The width of $f_0$ meson in the 
approximation of ``broad resonance'' depends on the product of coupling constants
 $g_{\phi KK}\cdot g_{fKK}$.  The $f_0$   fit parameters were  mass $m_f$,
  coupling constant $\frac{g^2_{fKK}}{4\pi}$ and the ratio of coupling constants
 $\frac{g^2_{fKK}}{g^2_{f\pi\pi}}$. The optimal fit parameters were obtained \cite{Ivanch}: 

\beq  m_f=971\pm 6 MeV, \ \Gamma_f=188^{+48}_{-33}MeV,
 \  \frac{g^2_{fKK}}{4\pi}=2.10^{0.88}_{0.56} GeV^2,
 \   \frac{g^2_{fKK}}{g^2_{f\pi\pi}}=4.1\pm 0.9.
 \label{pp6} \eeq 

The statistical accuracy did not allow to define the contribution  of $\sigma$ in the fit, so in (\ref{pp6})  
$\sigma$ was excluded  from the fit.

   The $\eta\pi^0$ mass spectrum was fitted also by the formulae  from  \cite{Pred}, but because of lower  
statistics  the ratio of  coupling constants was fixed  $\frac{g_{a\eta\pi}}{g_{aKK}}=0.85$ \cite{Pred}.
 The following $a_0$ optimal parameters were obtained:

\beq  m_a=992^{+22}_{-7} MeV,   \  \frac{g^2_{aKK}}{4\pi}=1.09^{+0.33}_{-0.24} GeV^2
 \label{etp4} \eeq 
 
The obtained value of  $a_0$  mass does not contradict to the PDG Table  value.  If the $a_0$  mass is fixed, 
one could obtain more accurate  value of coupling constant:

\beq    \frac{g^2_{aKK}}{4\pi} =0.83\pm  0.13  GeV^2
 \label{etp4} \eeq 

\begin{figure}[htb]   
\begin{minipage}[h]{70mm}
\epsfig{file=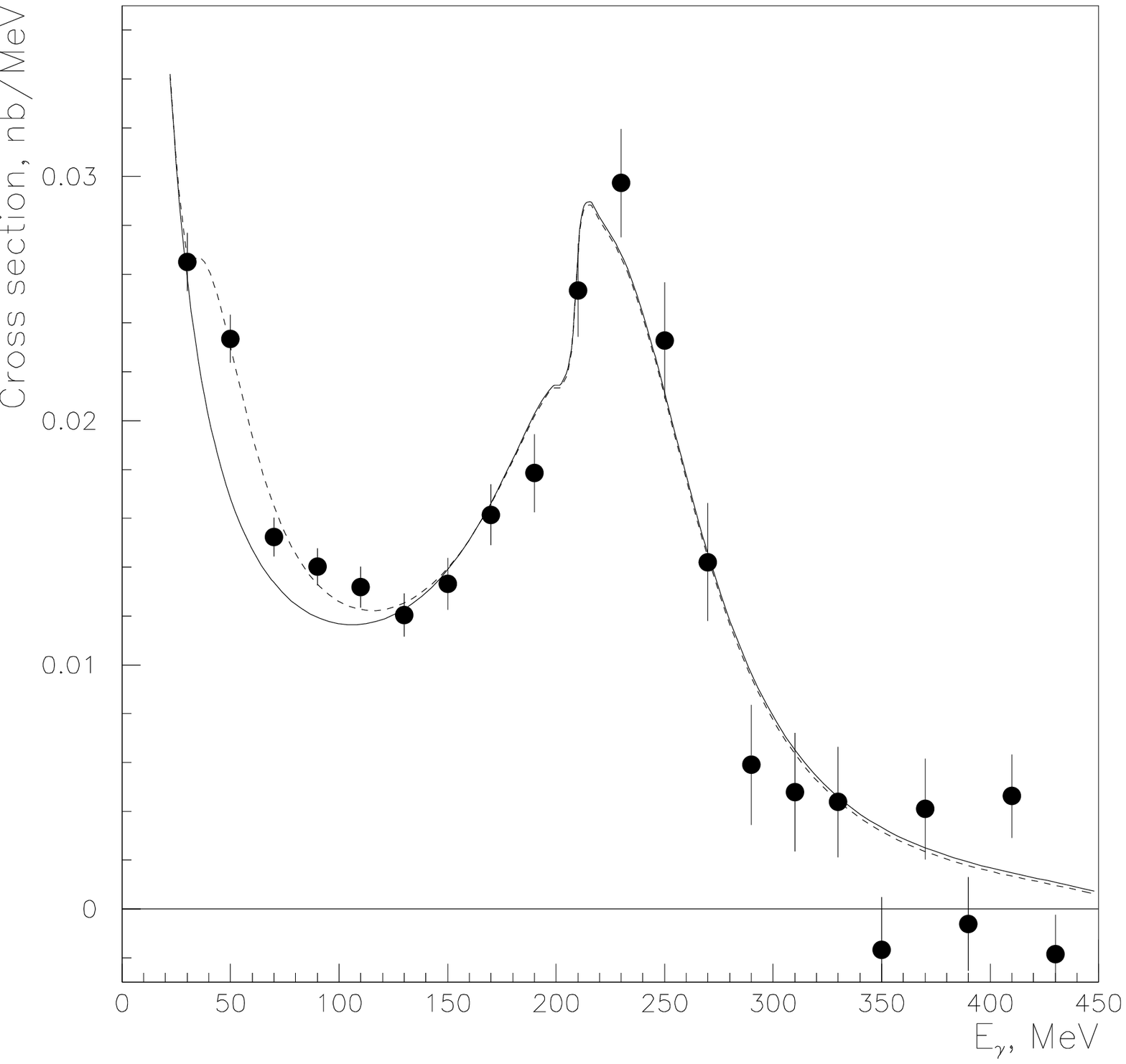, width=0.9\textwidth}
\caption{Recoil photon  spectrum in  the process
  $e^+e^-\to\pi^+\pi^-\gamma$ }
\label{rec}
\end{minipage}
\begin{minipage}[h]{70mm}
\epsfig{figure=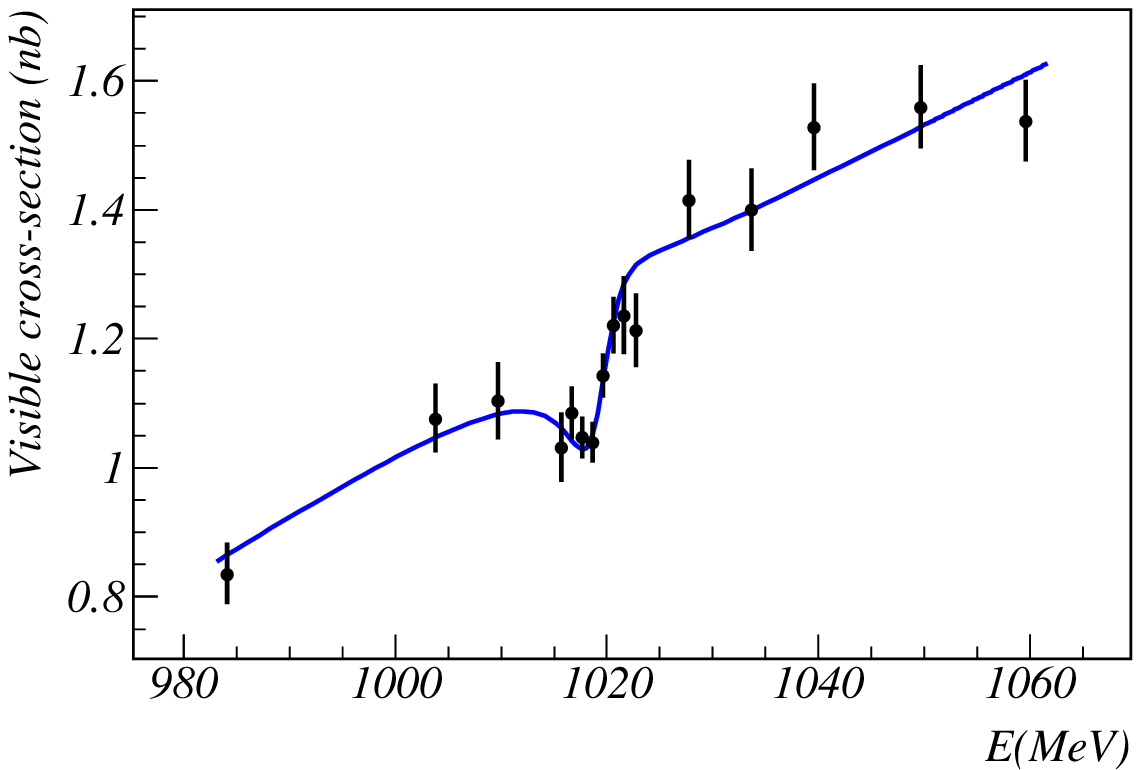,width=0.9\textwidth}
\caption{The detection cross section of the process
   $e^+e^-\to\omega\pi^0$ near $\phi$ meson}
\label{ompi}
\end{minipage}
\end{figure}

     CMD-2  carried out the search for the decay  $\phi\to\pi^+\pi^-\gamma$ in the reaction \cite{CMD2}: 
\beq e^+e^-\to\phi\to\pi^+\pi^-\gamma \label{pp7} \eeq
 with the goal to find a contribution of the $f_0\to\pi^+\pi^-$  channel in the final state.  On the contrary 
to the neutral  channel  $f_0\to\pi^0\pi^0$, there is a significant background 
from the nonresonant process $e^+e^-\to\rho\gamma\to\pi^+\pi^-\gamma$ 
and interference with the processes
  $e^+e^-\to\phi\to\pi^+\pi^-\gamma$ and $e^+e^-\to\rho\to\pi^+\pi^-\gamma$.
  It was found, that the process (\ref{pp7})  cross section energy dependence
 exhibits interference wave near point $2E=M_{\phi}$.  The recoil photon energy
 spectrum (fig.\ref{rec}) shows a peak at  $E_{\gamma}\simeq 220$~MeV 
due to the process  $e^+e^-\to\rho\gamma$ and  an enhancement at $E_{\gamma}\simeq 50$ MeV 
from the decay $f_0\to\pi^+\pi^-$,
which roughly corresponds to the mass difference between $\phi$ and $f_0$ mesons. To obtain
 the branching ratio  $B(\phi\to\pi^+\pi^-\gamma)$ the fitting of photon  spectra at different energy points 
was done using formulae from the work \cite{Gub}, which include contributions of the background reactions 
and the    $f_0\to\pi^+\pi^-$ decay.  The optimal value   of $f_0$ mass was $m_{f}=976\pm 5$~MeV,
  the  branching ratio:

\beq B(\phi\to f_0\gamma)=(1.93\pm 0.46\pm 0.59)\cdot 10^{-4} \label{pp8} \eeq 

\section{Discussion on the decays  $\phi\to f_0\gamma ,a_0\gamma$} In the list of new VEPP-2M data   
the results (7)-(12) are model  independent, because they are based on the total number of events.
 Other results - (13)-(18) use different assumptions, for instance,    $BR(\phi\to f_0\gamma)$  
(\ref{f4}),(\ref{f5}) is based on assumption of  $f_0$ dominance in the final $\pi^0\pi^0$ state. The main 
parameters of  $f_0$ and $a_0$ like their masses, widths, coupling constants were
 obtained from the description of these decays, proposed by N.Achasov  \cite{Pred}, so these parameters 
are also strongly model dependent.   Below we give a conclusion on the nature of  $f_0$ and $a_0$ 
scalars, which follows from the  model dependent data  (13)-(18)  and the  work \cite{Modl}.

        There are three main models, describing structure of   $f_0$  and $a_0$ scalars:
 $q\bar{q}$ model  ($n\bar{n}$ or  $s\bar{s}$),  molecular model
 $(K\bar{K})$ and 4-quark model $(q\bar{q}q\bar{q})$.   The general
 accepted opinion is that  $f_0$ and $a_0$ are difficult to fit into
 $q\bar{q}$ model.  This opinion is based on the existing experimental
 data.  For instance, the decays  $J/\psi\to f_0\gamma ,f_0\omega ,a_0\rho$
  are considerably suppressed in comparison with similar decays, where
 instead of  $f_0$ and $a_0$   the tensor mesons  $f_2$ or $a_2$ are
 produced. If   $f_0$ and $a_0$ are  $q\bar{q}$ mesons, their production
 in  $J/\psi$ decays should be of  the same order  as the production
 of tensor mesons.  

Another example is two-photon width  of  $f_0$ and $a_0$.
 The experimental value $\Gamma\simeq 0.3$ keV is smaller
 than the value 0.6 keV predicted in $K\bar{K}$-model and  value
 $0.6\div15. $ keV   in   $q\bar{q}$ model. But the four-quark
 model prediction (0.3 keV) agrees with experiment.
   
\begin{table}[htb]
\begin{center}
\begin{tabular}{l|l|l|r|r|r}  \hline
Model &  $n\bar{n}$ & $s\bar{s}$ & $K\bar{K}$ & $q\bar{q}s\bar{s}$ & Exp-t \\  \hline
$Br(\phi\to f_0\gamma)\cdot 10^4$ &
 0.45 & 0.55 & 0.1 & 2.5 & $3.0\pm 0.4$\\ \hline
$Br(\phi\to a_0\gamma)\cdot 10^4$ &
 0.25 & --- & 0.1 & 2.0 & $0.88\pm 0.13$\\ \hline
\end{tabular}
\caption{The comparison of VEPP-2M data with different  $f_0$ and $a_0$ models.}
\label{Tabl}
\end{center}
\end{table}

       The radiative decay $\phi\to f_0\gamma ,a_0\gamma$ measurements
  were long awaited as a new test of  $f_0$ and $a_0$ nature. The Table
 1 shows the comparison of different models predictions with averaged
 experimental data at VEPP-2M (see preceding section).   The accuracy
 of the model predictions \cite{Modl} is about $50\%$.  The conclusion
 from the Table 1 is that VEPP-2M data are in good agreement with four-quark 
model of   $f_0$ and $a_0$. But we remind the reader, that experimental
 data in the Table 1 assume, that  $f_0$ and $a_0$ dominate in the final state
 of the reactions  (1) and (2). This assumption is in good agreement with 
experimental spectra, but the  present accuracy is not sufficient  to exclude
contributions of other scalars  into the final state. 

      There is one remark, concerning  the decay  $\phi\to a_0\gamma$. Its
 branching ratio is  close to that of $\phi\to\eta '\gamma$. So, $a_0$ should 
contain strange quarks like $\eta '$, which is impossible for $q\bar{q}$ isovector meson, but
is quite natural if  $a_0$ is a four quark $q\bar{q}s\bar{s}$ meson.
  
     This discussion was based mainly on the work \cite{Modl}, where detailed
 analysis of existing data for    $f_0$ and $a_0$ mesons, regarding their nature,
  is given.

\section{Other rare $\phi$ decays}
\begin{figure}[htb] 
\begin{minipage}[h]{75 mm}
\epsfig{figure=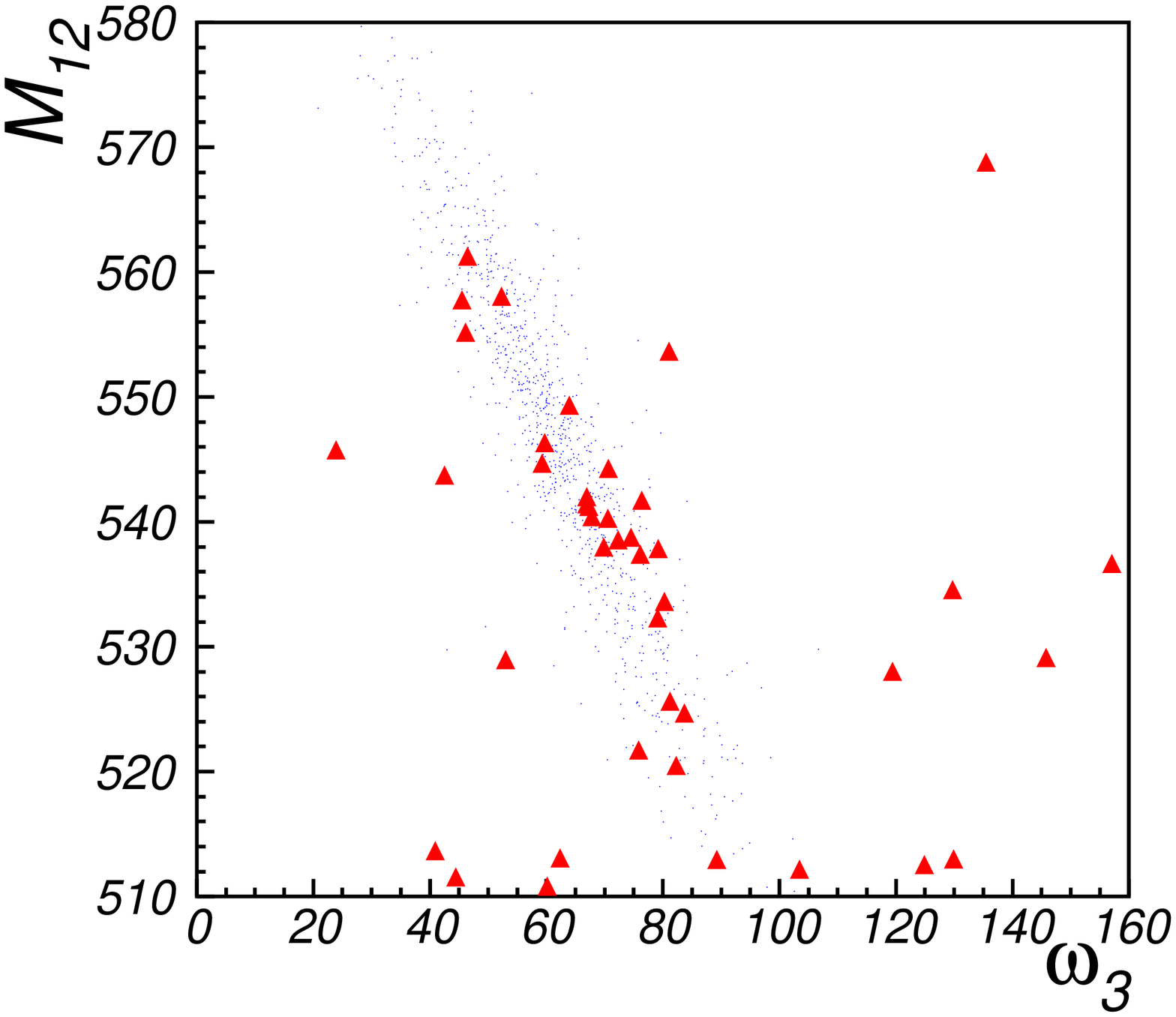, width=0.9\textwidth}
\caption{Two-photon invariant mass in $\eta\to\gamma\gamma$ decay vs. recoil
 photon energy in a search for 
$\phi\to\eta '\gamma , \eta '\to\eta\pi^+\pi^- , \eta\to\gamma\gamma$ decay.}
\label{eta'}
\end{minipage}
\hspace{\fill}
\begin{minipage}[h]{75  mm}
\epsfig{figure=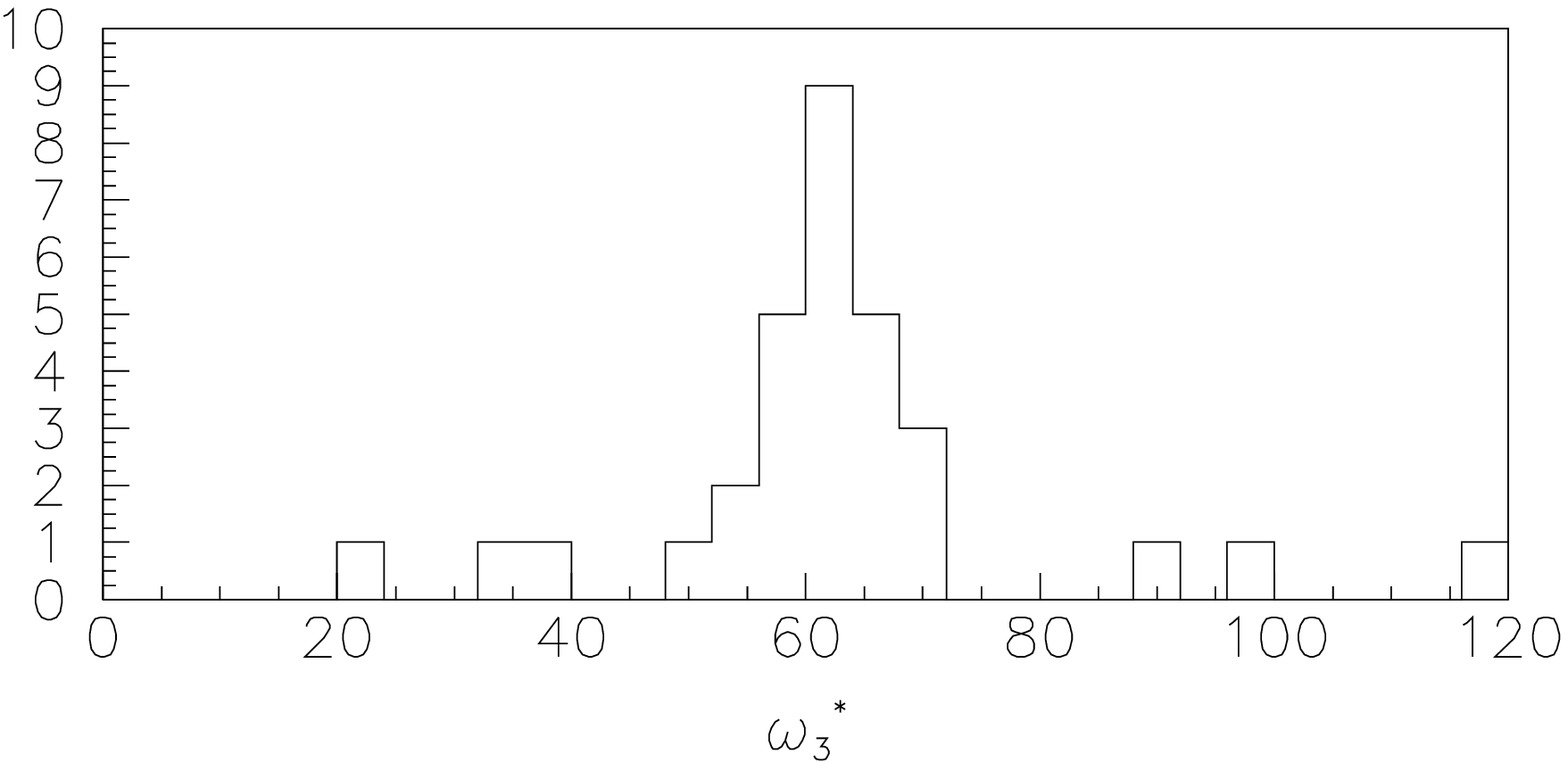, width=0.9\textwidth}
\caption{Recoil photon energy spectrum in the $\phi\to\eta '\gamma$ decay.}
\label{ETA'} 
\end{minipage}
\end{figure}

   The large number of produced $\phi$ mesons at both SND and CMD-2 detectors
  ($N_{\phi}\simeq 4.5\cdot 10^{7}$)   allows to carry out the search of  rare
 $\phi$ decay.  The long awaited decay $\phi\to\eta '(958)\gamma$ was first observed
 by CMD-2 \cite{CMD3}. In the decay chain
 $\phi\to\eta '\gamma ,     \eta '\to\eta\pi^+\pi^-,  \eta\to\gamma\gamma$
 the branching ratio was $(8.2^{+2.1}_{-1.9})\cdot 10^{-5}$.  
For another chain  $\eta '\to\pi^+\pi^-\pi^0(\gamma )$  the CMD-2 result
 was $(5.8\pm 1.8)\cdot 10^{-5}$  \cite{FED}. Later SND confirmed the existence
 of the decay $\phi\to\eta '\gamma$ with the branching ratio of $6.7^{+3.4}_{-2.9} \cdot 10^{-5}$
     \cite{ETG'}.   The clear signature of the decay  $\phi\to\eta '\gamma$ is
 demonstrated in figs.\ref{eta'}, \ref{ETA'}. The averaged  value of branching
 ratio is $BR(\phi\to\eta '\gamma) = (6.9\pm 1.2)\cdot 10^{-5}$. The statistical
 significance is greater than 5 standard deviations.
 This result is in agreement with nonrelativistic
 quark model prediction of $(6\div10)\cdot 10^{-5}$ \cite{NQM}.  At present level
 of the accuracy no significant admixture of gluonim in  $\eta '$ is seen.

     The $\phi\to\pi^+\pi^- ,\omega\pi^0$ decays are double suppressed - by isospin invariance
 and OZI rule.  The $\phi\to\pi^+\pi^-$ decay was already observed.  Its PDG Table
 value  is $Br (\phi\to\pi^+\pi^-)=(0.8^{+0.5}_{-0.4})\cdot 10^{-4}$, while the second
 decay  $\phi\to\omega\pi^0$ was not observed yet. SND performed  the search for this
 decay in the the reaction \cite{Ompi}: 
\beq e^+e^-\to\omega\pi^0\to\pi^+\pi^-\pi^0\pi^0   \label{omegapi} \eeq
The clear interference pattern in energy dependence  of the process (\ref{omegapi})
 was observed (fig.\ref{ompi}). The decay amplitudes and branching ratios are the
 following \cite{H99}:
\beq Re(Z)= 0.112\pm 0.015, \ Im(Z) = -0.104\pm 0.022,  \
  Br(\phi\to\omega\pi^0)=(4.6\pm 1.2)\cdot 10^{-5} \label{ompres} \eeq
The theoretical  prediction \cite {KOZH} for the branching ratio is about
 twice larger. In our case the  real part Re(Z)  is too low.  The observed
 disagreement could be due to existence of direct  $\phi\to\omega\pi^0$
 transition or nonstandard mixing of light vector mesons. 

    In similar way   was studied the cross section of the process 
\beq e^+e^-\to\pi^+\pi^-  \label{pipi} \eeq
The results of fitting are \cite{H99}:
\beq Re(Z)= 0.061\pm 0.005, \ Im(Z) = -0.042\pm 0.006,  \
  Br(\phi\to\pi^+\pi^-)=(7.1\pm 1.0\pm 1.0)\cdot 10^{-5} \label{pipires} \eeq

   The accuracy of the measurement is about 3 times higher than in PDG Tables.
 But here again SND result for real part  Re(Z) is lower than predicted in
 \cite{KOZH} and preliminary result of CMD-2 \cite{PPCMD}:
\beq   Br(\phi\to\pi^+\pi^-)=(18.1\pm 2.5\pm 1.9)\cdot 10^{-5} \label{pipicmd} \eeq
The disagreement between CMD-2 and SND in  $Br(\phi\to\pi^+\pi^-)$
 is 3 standard deviations.

     Both detectors studied the rare decay $\phi\to\mu^+\mu^-$.
  The result of SND is \cite{mumu}
\beq   Br(\phi\to\mu^+\mu^-)=(33.0\pm 4.5\pm 3.2)\cdot 10^{-5} \label{musnd} \eeq
The result of  CMD-2 is \cite{PPCMD}
\beq   Br(\phi\to\mu^+\mu^-)=(28.0\pm 3.0\pm 4.6)\cdot 10^{-5} \label{mucmd} \eeq
     The full review of other $\phi$ rare decay studied  at VEPP-2M can be found in refs \cite{PPCMD, SNDATA}.

\section{Decays  $\rho ,\omega\to\pi^0\pi^0\gamma$ }  The decays
  $\rho ,\omega\to\pi^0\pi^0\gamma$ are of interest for the study
 of the possible low-mass scalar resonance $\sigma$, decaying into
 $\pi\pi$ final state.  Some contributions are expected also from the
 $\rho ,\omega\to\omega\pi^0 ,\rho\pi \to\pi^0\pi^0\gamma$ decays.
   In our work  \cite{NData}, where the $\rho\to\pi^+\pi^-\gamma$ decay
 was studied, an enhancement was observed in the high end of the photon
 bremsstrahlung spectrum, which can be interpreted as a manifestation of a light bound state,
 possibly    $\sigma$ resonance. Later in Protvino, the decay
  $\omega\to\pi^0\pi^0\gamma$ was observed with the branching ratio
 $(7.2\pm 2.5)\cdot 10^{-5}$, which is $\sim 3$ times larger, than
 expected in Vector Dominance Model (VDM). 
\begin{figure}[htb]
\begin{minipage}[h]{70mm}
\epsfig{figure=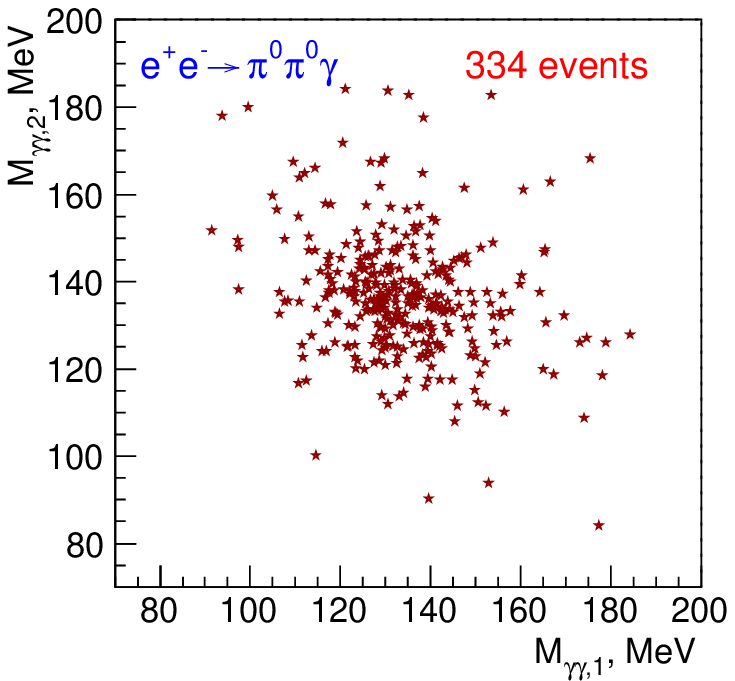,width=0.9\textwidth}
\caption{2-dim. scatter plot of the best neutral pion candidates
 in the search for the process $e^+e^-\to\pi^0\pi^0\gamma$. }
\label{2p0}
\end{minipage}
\begin{minipage}[h]{70mm}
\epsfig{figure=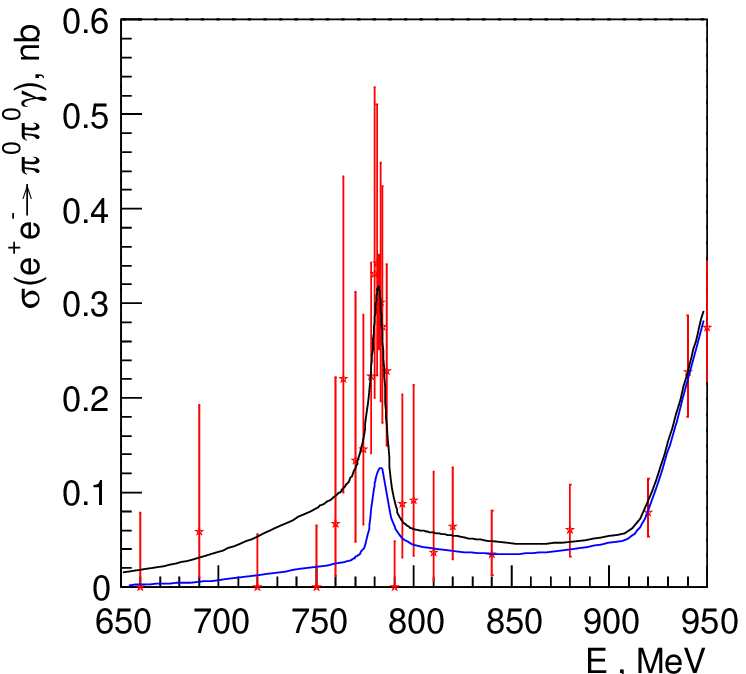,width=0.90\textwidth}
\caption{Born cross section of the process $e^+e^-\to\pi^0\pi^0\gamma$;
 upper curve is a fit, lower curve - VDM prediction.  }
\label{p0p0g}
\end{minipage}
\end{figure}

    In the recent work \cite{H99} we studied  neutral final state in the reaction  
$e^+e^-\to\rho ,\omega\to\pi^0\pi^0\gamma\to 5\gamma$. Fig. \ref{2p0} shows a 2-dimensional
 plot of the best neutral  pion candidates, found in 5-photon final state.
  The  measured cross section was fitted by the sum of the Breit--Wigner 
contributions from $\omega$ and $\rho$ resonances.  The Born cross section and fitting
curves are shown on the fig.\ref{p0p0g}. One can see, that the measured cross
 section considerably exceeds  VDM prediction. The fit parameters are the following:

\beq  BR(\omega\to\pi^0\pi^0\gamma)=(8.4^{+4.9}_{-3.1}\pm 3.5)\cdot 10^{-5},
 \  \Gamma_{\omega\pi^0\pi^0\gamma} \simeq 0.7 keV  \label{wppg} \eeq
\beq
BR(\rho\to\pi^0\pi^0\gamma)=(4.2^{+2.9}_{-2.0}\pm 1.0)\cdot 10^{-5},
 \  \Gamma_{\rho\pi^0\pi^0\gamma}\simeq 6 keV, (without \  \omega\pi^0 ) \label{rppg}
\eeq
So, the result (\ref{wppg})  confirms the PDG value of  $BR(\omega\to\pi^0\pi^0\gamma)$.
  Both branching ratios (\ref{wppg})  and  (\ref{rppg})  are considerably ($\sim 4$ times)
 higher than VDM estimates. The possible explanation of this enhancement could be 
a contribution  of light scalar $\sigma$, decaying into $\pi^0\pi^0$.
  It was suggested  by Jaffe \cite{Jaf}, that  $\sigma$ could be lightest member of the four-quark nonet
 with the structure $u\bar{u}d\bar{d}$. Because of superallowed 
 $\sigma\to\pi\pi$ decay,  $\sigma$ is very broad. Among other members of four-quark
 nonet there  are $f_0(980)$ and  $a_0(980)$ - the particles with also superallowed
 but phase space suppressed decay into $K\bar{K}$.  So, both $f_0(980)$ and
  $a_0(980)$ have a narrow width 50--100~MeV.    The further investigation of the decays
$\phi ,\rho ,\omega\to\pi^0\pi^0\gamma$ and in particular study of the $\pi^0\pi^0$
 decay mass spectra could clarify the nature of light scalar mesons.

\section{The process $e^+e^-\to\pi^+\pi^-\pi^0$ above $\phi$ resonance}

\begin{figure}[htb]
\begin{minipage}[h]{70mm}
\epsfig{figure=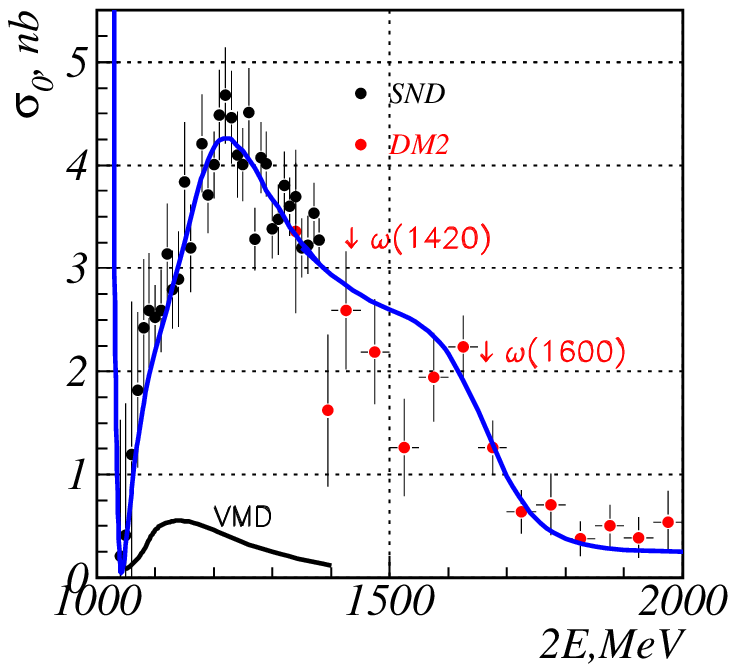,width=0.95\textwidth}
\caption{Born cross section of the process  $e^+e^-\to\pi^+\pi^-\pi^0$ (linear scale)}
\label{pi3}
\end{minipage}
\begin{minipage}[h]{70mm}
\epsfig{figure=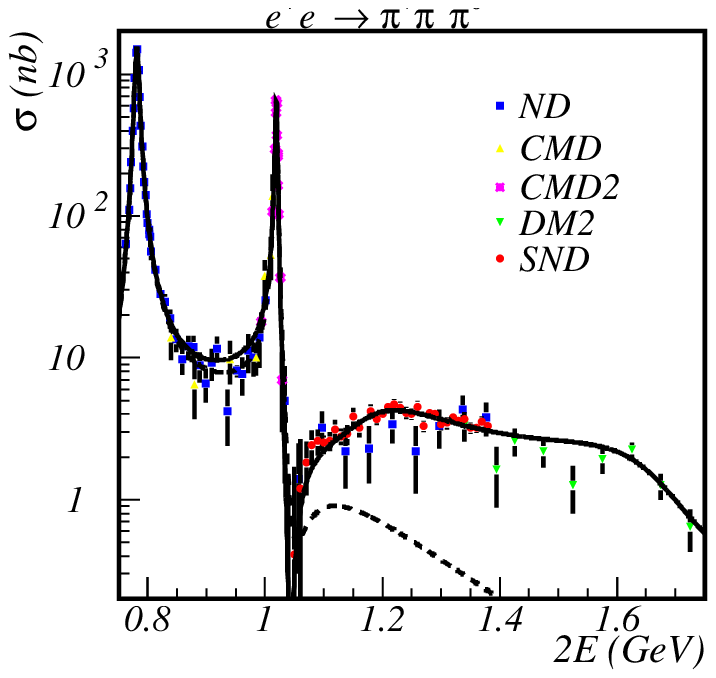,width=0.95\textwidth}
\caption{Born cross section of the process  $e^+e^-\to\pi^+\pi^-\pi^0$ (logarithmic scale)}
\label{3pi_log}
\end{minipage}
\end{figure}
   The energy region above $\phi$ was scanned with the goal to measure
 $e^+e^-$- annihilation cross sections and quantity R - the ratio of
 total hadronic cross section to muon pair production cross section.  Among
 the processes under study the process 
                 \beq e^+e^-\to\pi^+\pi^-\pi^0  \label{3pi} \eeq
is of particular interest, because earlier  it was measured with poor
 accuracy and new possible  isoscalar vector resonances could be found here.
  The study of the process (\ref{3pi}) was done by SND detector in the
 energy range 2E=1.04--1.38 GeV \cite{p3plt} .  

  The measured cross section, shown in fig.\ref{pi3},\ref{3pi_log} is 
in agreement with previous data from ND experiment \cite{NData} and well
 matches DM2 measurements at higher energies \cite{DM2}. The systematic
 error in the cross section  is $\sim 10\%$, but it grows up to $ 50\%$  closer
  to $\phi$  because of radiative corrections. The Born cross section 
 in fig.\ref{pi3}  shows a broad peak with the visible position at 
 $2E\simeq 1200$~MeV. To describe the cross section in terms of sum 
of vector mesons,  the fit was done including $\omega (783)$,
  $\phi (1020)$,  $\omega (1600)$ and an additional  $\omega$-like state,
 named  $\omega (1200)$, with its mass and width set free.  For two latter
 resonances the widths were assumed independent of energy.  The  optimal
 fit parameters strongly depended  on interference phases choice. The
 best fit occurs at the following phase set: $\phi_{\omega (783)}=0$,
   $\phi_{\phi (1020)}=\pi$, 
$\phi_{\omega (1200)}=\pi$,  $\phi_{\omega (1600)}=0$.  The  $\omega (1200)$
  parameters are :

\beq M_{eff}=1170\pm 10 MeV, \ \Gamma_{eff}=187\pm 15~MeV, \sigma_{max}=7.8\pm 1.0 nb, 
\eeq
 The parameters of the resonance  $\omega (1600)$ are confirmed by the fit,
 but another  resonance $\omega (1420)$  is not seen in our fit. If the
 existence of $\omega (1200)$ is  confirmed, the question of its nature arises.
 It could be either first radial excitation $2^3S_1$ or radial excitation (D-wave)
    $1^3D_1$ of $\omega (783)$.  In any case, new analyses of isoscalar cross
 sections data  are needed to clarify the problem of   $\omega$ family excitations.    

  \section{Project VEPP-2000 } 
\begin{figure}[htb]
\begin{minipage}[h]{150mm}
\epsfig{figure=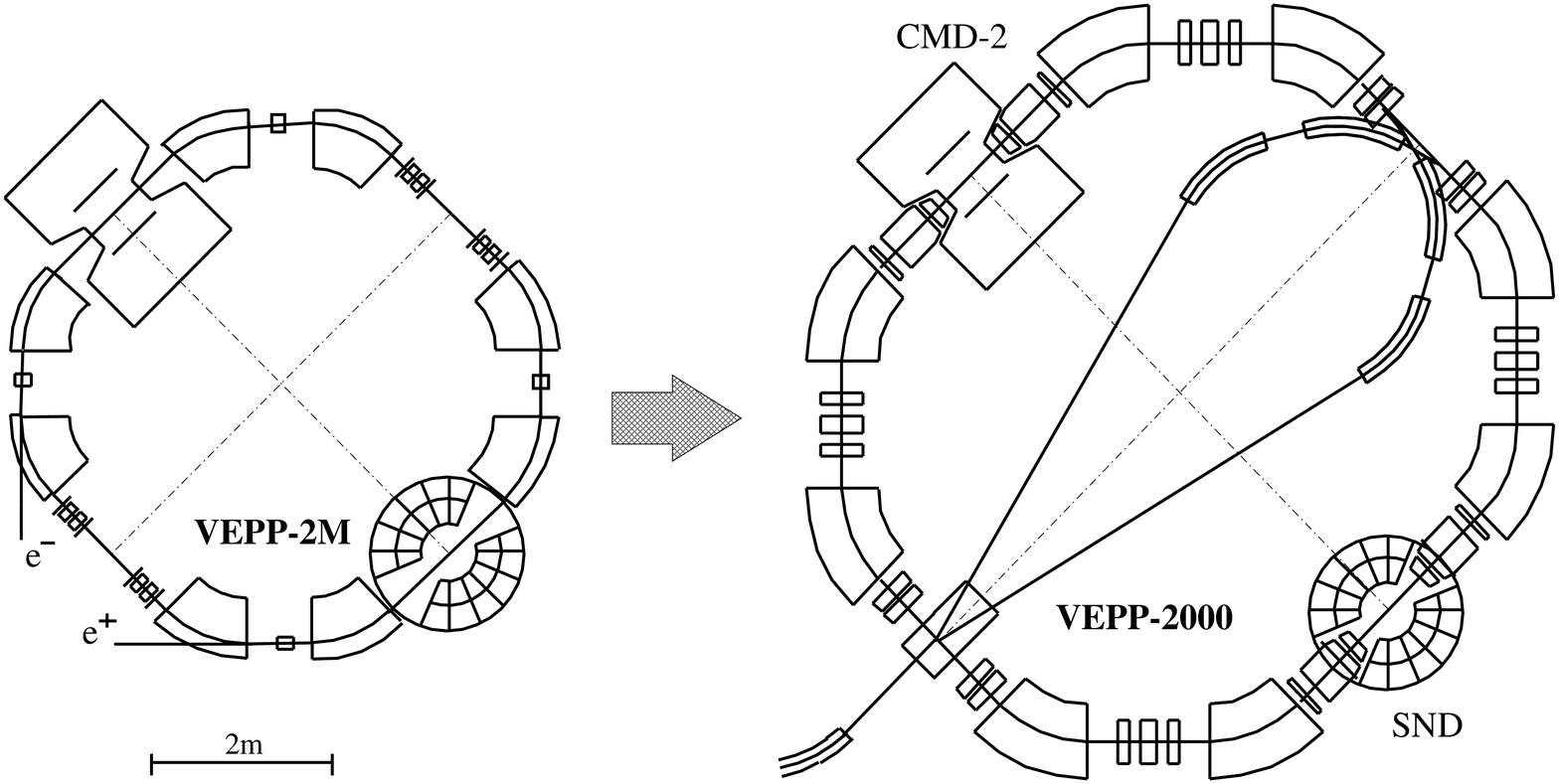, width=0.9\textwidth}
\caption{VEPP-2000 project.  Left chart depicts existing VEPP-2M ring.
 The right one is the newly proposed  VEPP-2000 ring.  Two collider
 detectors CMD-2 and SND, located opposite each other, are shown as well. }
\label{VEPP}
\end{minipage}
\end{figure}

A new project is studied now in Novosibirsk.
 It is planned to replace VEPP-2M ring which has a maximum center of mass energy of $2E=$1400~MeV
  by a new one with the higher energy up to $2E=$2000~MeV.   Fig. \ref{VEPP} shows the
 location of the  new and the old rings in the VEPP-2M hall. A remarkable  feature
 of the new collider is a round beam optics, where instead of conventional quadrupole lenses
 the  superconductive solenoids are used.  The beam itself has equal horizontal and
 vertical size, which promises the higher luminosity in single bunch mode.  The future
  collider is named VEPP-2000. Its designed luminosity is $10^{32} cm^{-2}s^{-1}$  at $2E=$2000~MeV
 and  $10^{31} cm^{-2}s^{-1}$ at $2E=$1000~MeV. 
       
     The design and construction of VEPP-2000 is planned to start in 2000. The physical
 program is aimed to detailed study of $e^+e^-$ annihilation processes in the energy
 range $2E=$1--2~GeV. 
\section{Evidence of possible exotic baryon X(2000)} 
 Among the contributed papers, there is one, presented by L.Landsberg,
 IHEP, Protvino \cite{Contr},   related to the subject of exotic hadrons.
 In this work the diffractive production of baryon  resonances was studied with the
 SPHINX setup in the reaction: 
\beq p+N\to YK+N ,  \ \ Y= [\Sigma^0K^+], \ \ \Sigma^0\to\Lambda\gamma  \label{LG1}  \eeq  

The mass spectrum of $\Sigma^0K^+$ shows a clear peak at $\simeq 2000$~MeV,
 which is referred below as X(2000).  The fitting gives more accurate values:
 $M_X=1989\pm 6 MeV, \ \Gamma_X=91\pm 20 MeV$. The statistical significance
 is more than 10 standard deviations.
  The production cross section  is $95\pm 20$ nb. The
 unusual dynamic properties of X(2000) are the following:  1 - the value
 $R=\frac{Br(X\to\Sigma K)}{Br(X\to nonstrange)}\geq 1$,  while for usual $qqq$
 isobar $R\sim 10^{-2}$;  2 - the width of X(2000)  $\Gamma_X\leq100$~MeV,
 which is considerably less than for isobars -  $300\div400$~MeV.
    
     All these properties of X(2000) allow to consider it as a  serious candidate
 for pentaquark exotic baryon with hidden strangeness $uuds\bar{s}$.  Latest data
 from SPHINX experiment  confirmed the existence of X(2000) in another final state
$Y= [\Sigma^+K^0], \   \Sigma\to p\pi^0, \ K^0\to\pi^+\pi^-$. 
 New preliminary data from SELEX experiment at Fermilab also supports X(2000).
 In analysis of the reaction  $\Sigma^-+N\to\Sigma^-K^+K^-+N$            
they observed a peak in   $Y=[\Sigma^-K^+]$ system,  with parameters close 
to X(2000) : $M_X=1962\pm 12 MeV, \ \Gamma_X=96\pm 32 MeV$.

       Now a lot of statistics is accumulated on tapes with upgraded SPHINX detector.
 The analysis of new data is in progress. 

\section{General Conclusions}

\begin{itemize}
\item 
Experiments were carried out in Novosibirsk at VEPP-2M $e^+e^-$ collider
with two detectors SND and CMD-2 with total integrated luminosity
$\simeq 50 pb^{-1}$ and total number of produced $\phi$ mesons $\sim 4\cdot
10^7$.
\item 
Electric dipole radiative decays $\phi\to\pi\pi\gamma$, $\eta\pi^0\gamma$
were observed with branching ratios $\sim 10^{-4}$,
indicating exotic 4-quark structure of lightest scalars 
$f_0(980), a_0(980)$.
\item 
Several new rare  $\phi$-meson decays were observed with branching fractions
$\sim 10^{-4}\div 10^{-5}$, e.g., $\phi\to\omega\pi^0$, $\phi\to\eta'\gamma$, 
$\phi\to 4\pi$,   $\phi\to\pi^0e^+e^-$,\ldots
\item 
A resonance-like structure in  $e^+e^-\to\pi^+\pi^-\pi^0$ cross section near
$2E\simeq$1.2 GeV was observed, which might be a manifestation of the lightest
excited $\omega$ state,
\item 
 The decays  $\rho,\omega\to\pi^0\pi^0
\gamma$ were seen. Their rates exceed VMD level, which might be a manifestation
of lightest scalar state $\sigma$(400-1200), decaying into $\pi^0\pi^0$.
\item 
Design and construction of a new  VEPP-2000  $e^+e^-$ machine with round 
beams  to replace existing VEPP-2M ring are started in 
Novosibirsk. The maximum  designed energy of the new machine is $2E$=2000~MeV, 
designed luminosity --- $L=1\cdot 10^{32}$.
\item 
 In SPHINX experiment, Protvino, a narrow X(2000) state with a width $\Gamma\simeq$90~MeV
 was observed. It is proposed as a candidate for pentaquark exotic baryon $qqqs\bar{s}$
with hidden strangeness. 
\end{itemize}
\bigskip
     The author is grateful to Nikolai Achasov, Vladimir Golubev and Evgeny Solodov
 for numerous fruitful discussions.

\def\Discussion{
\setlength{\parskip}{0.3cm}\setlength{\parindent}{0.0cm}
\bigskip\bigskip      {\Large {\bf Discussion}} \bigskip}
\def\speaker#1{{\bf #1:}\ }

\Discussion

\speaker{L. G. Landsberg (IHEP, Protvino)}
What can you say about the two-photon production of the $a_0$ and $f_0$ mesons? 
Are these data in agreement with the $qq\bar q\bar q$ or
 $q\bar q$ models for these mesons?   

\speaker{Serednyakov} The measured two photon widths
 of $a_0$ and $f_0$ $\simeq 0.3$ KeV  are significantly lower
 than predictions of $q\bar q$ model and agree with  $qq\bar q\bar q$ model.

\speaker{Norbert Wermes (Bonn University)}
Are the two detectors at VEPP-2M capable of  measuring  $R_{\rm had}$? Will
they be able to perform  a scan in energy?

\speaker{Serednyakov}
 Both detectors have already accumulated $50\, pb^{-1}$ of data  and continue data taking in
the energy range from 0,4 to 1.4 GeV.  The measurement of $R$ is one of the major goals
of these experiments.

\speaker{B.F.L. Ward (University of Tennessee)}
In your table of model predictions vs. experiment, why do you say that value 2.5, for the
$q\bar q$ model, is farther than 20, for the 4-quark model, from the
experimental value of 9?

\speaker{Serednyakov} Because the accuracy of theoretical prediction is
 about $40\div50\%$ , so the 4-quark value $20\pm 10$  considerably better
 agrees with experimental value  9  than 2-quark value $2.5\pm 1.3$. 

\speaker{Harry Lipkin (Weizmann Institute)}
There are very beautiful data on $D_s \rightarrow f^0\pi \rightarrow 3\pi$
  from  Fermilab and on $\bar p p \rightarrow f^-\pi \rightarrow 3\pi$ from 
 CERN.  Dalitz  plot analyses of these reactions should be available soon.

\speaker{Serednyakov}
  The data on $D_s\rightarrow f^0\pi$ decay  show that $f_0$ should include
 $s$-quarks.  The $n\bar n$ structure of $f_0$ is not supported in 
 $D_s \rightarrow f^0\pi$ decay.
 
\end{document}